\documentclass[conference]{IEEEtran}
\IEEEoverridecommandlockouts
 \usepackage{lipsum,graphicx,multicol}
\usepackage{color,soul}
\usepackage{mathtools}
\usepackage{graphicx}
\usepackage{color,soul}
\usepackage{setspace}
\usepackage{multirow}
\usepackage{tablefootnote}
\usepackage[norelsize, linesnumbered, ruled, lined, boxed, commentsnumbered]{algorithm2e}
\usepackage[normalem]{ulem}
\usepackage[font=small,labelfont=bf]{caption}
\usepackage{subcaption}
\usepackage{bm, amsmath, amssymb, amsthm}
\usepackage{amsfonts}
\usepackage {amssymb,amsmath,xcolor}
\usepackage{booktabs}
\usepackage{cite}
\usepackage{array}
\usepackage[acronym]{glossaries}
\usepackage{cases}
\usepackage{cleveref}

\def\BibTeX{{\rm B\kern-.05em{\sc i\kern-.025em b}\kern-.08em
    T\kern-.1667em\lower.7ex\hbox{E}\kern-.125emX}}
    
  \newtheorem{theorem}{Theorem}

\newtheorem{corollary}{Corollary}
\newtheorem{proposition}{Proposition}
\newtheorem{lemma}{Lemma}

\newcommand{\bieee}{\begin{IEEEeqnarray}{rCl}}
\newcommand{\eieee}{\end{IEEEeqnarray}}

\SetKwInput{KwInput}{Input}                
\SetKwInput{KwOutput}{Output}              

\begin{document}

\title{On Securing Analog Lagrange Coded Computing from Colluding Adversaries
}

\author{\IEEEauthorblockN{Rimpi Borah and J. Harshan}
\IEEEauthorblockA{Department of Electrical Engineering, Indian Institute of Technology Delhi, India}}

\maketitle

\begin{abstract}
Analog Lagrange Coded Computing (ALCC) is a recently proposed coded computing paradigm wherein certain computations over analog datasets can be efficiently performed using distributed worker nodes through floating point implementation. While ALCC is known to preserve privacy of data from the workers, it is not resilient to adversarial workers that return erroneous computation results. Pointing at this security vulnerability, we focus on securing ALCC from a wide range of non-colluding and colluding adversarial workers. As a foundational step, we make use of error-correction algorithms for Discrete Fourier Transform (DFT) codes to build novel algorithms to nullify the erroneous computations returned from the adversaries. Furthermore, when such a robust ALCC is implemented in practical settings, we show that the presence of precision errors in the system can be exploited by the adversaries to propose novel colluding attacks to degrade the computation accuracy. As the main takeaway, we prove a counter-intuitive result that not all the adversaries should inject noise in their computations in order to optimally degrade the accuracy of the ALCC framework. This is the first work of its kind to address the vulnerability of ALCC against colluding adversaries.     
\end{abstract}


\section{Introduction}

As the primary objective of cloud based computing architectures is to facilitate efficient execution of large computational tasks, the design focus is typically on realizing distributed computation features. However, security objectives such as privacy, integrity and availability are provided at an independent layer through off-the-shelf crypto-primitives. As a promising alternative to the above mentioned layered architecture, secure coded computing architectures have been recently studied wherein coding- and information-theoretic tools have been used to provide efficient computations \cite{new2} in the presence of \emph{stragglers} \cite{b11}, \emph{honest-but-curious} workers \cite{b13}, and \emph{adversarial} workers \cite{b12}. For instance, Lagrange Coded Computing (LCC) \cite{b1}, \cite{new1}  is one such distributed coded computation framework that provides all the above mentioned features. Although LCC is effective, its computation accuracy takes a hit due to overflows arising from finite field operations, when the dataset is large.

To circumvent the accuracy issues of LCC, Analog Lagrange Coded Computing (ALCC) \cite{b2} has been proposed to directly work on analog datasets through floating point representation. Although the ALCC framework provides privacy on the datasets from honest-but-curious workers, it is not known if it is robust to the presence of a few adversarial workers. In this context, adversarial workers refer to those workers in the distributed setup that intentionally return erroneous computations. Inspired by this research gap, in this work, we study the security vulnerabilities of ALCC against adversarial workers, and propose novel methods to enhance their computational accuracy in the presence of adversaries.

Our preliminary investigation reveals that the vanilla ALCC provides poor accuracy in the presence of adversarial workers. Towards fixing this issue, we identify that in the absence of adversaries, the computations returned by the workers in ALCC are essentially a set of codewords of a Discrete Fourier Transform (DFT) code \cite{b3, ShH}. As a consequence, if a few workers return noisy computations, the introduced errors can be corrected by applying the error-correction algorithms for the DFT code before the interpolation step of ALCC. Thus, the existence of DFT decoders answers the question posed in \cite{b2}. Besides the above observation, we show that the structure of the computations returned from the workers compels the master node of ALCC to solve the roots of a large set of error-locator polynomials (of varying degrees) which may have a common set of solutions. Identifying this ALCC requirement, we propose a novel variant of error localization to jointly solve the roots of multiple error-locator polynomials. Furthermore, when such DFT based decoders are implemented on systems with precision errors, we prove that the proposed joint localization provides significant improvement in accuracy when compared to using off-the-shelf DFT decoders \cite{b3}. While the proposed ALCC variant offers promising accuracy in the presence of a few adversarial workers, we also show that the presence of precision errors in practical ALCC settings could be exploited by the adversaries to degrade its accuracy. In particular, backed by analytical results on the error rates of the proposed DFT decoders, we propose two classes of colluding attacks based on the communication-overhead restrictions imposed by their colluding channels. Through analytical results, we prove that the optimal strategy for the adversaries is to ensure that not all adversaries inject noise on their computations. Experimental results are also presented to justify our theoretical claims on the proposed attacks. 

Due to space constraints, we have omitted the proofs of some propositions, lemmas and theorems. 


\section{Background on ALCC}
\label{SMPS}
The ALCC set up consists of a master node which is connected to $N$ workers via dedicated links. Since the vanilla ALCC framework in \cite{b2} involves some stragglers and honest-but-curious workers, we will give a background assuming their presence. Let $\mathbf{X}=(\mathbf{X}_1,\ldots,\mathbf{X}_k)$ be the dataset where $\mathbf{X}_i \in \mathbb{R}^{m\times n} $ for $ i\in[k]$ where $[k]$  denotes \{$1,2,\ldots,k$\}. The main objective of ALCC is to evaluate a polynomial $f:\mathbb{R}^{m\times n}$ $\rightarrow$ $\mathbb{R}^{u\times h}$ over the dataset $\mathbf{X}$ in a decentralized fashion while guaranteeing the resiliency from a certain number of stragglers indicated by $s$ and privacy from at most $t$ honest-but-curious workers. Furthermore, the scope of computation is such that $f$ is a $D$-degree polynomial function wherein all the entries of the output matrix are multivariate polynomial functions of the entries of the input with a total degree at most $D$. For instance, if $\mathbf{X} \in \mathbb{R}^{m\times n}$ is such that $x_{pq}$ is the $(p,q)$-th entry of $\mathbf{X}$, for $p \in [m]$ and $ q\in [n]$, then $\mathbf{Y} = f(\mathbf{X})$ refers to the output of computation such that $y_{ij}$, which is the $(i,j)$-th entry of $\mathbf{Y}$, for $i\in [u]$, $j\in [h]$ is of the form $y_{ij} =f_{ij}(x_{11},x_{12}, \ldots, x_{mn})$, where each $f_{ij}$  is a multivariate polynomial of degree $D$. In the following sections, we explain a distributed way to compute $\mathbf{Y} = f(\mathbf{X})$ when the function $f$ is known to all the workers. 


\subsection{Encoding in ALCC}
This section describes a way to encode the data set $\mathbf{X}$ into shares in order to distribute them among the workers. Let $\mathbf{Z}=(\mathbf{X}_1,\mathbf{X}_2,\ldots,\mathbf{X}_N,\mathbf{N}_1,\mathbf{N}_2,\ldots,\mathbf{N}_t)$ where $\{\mathbf{N}_1, \mathbf{N}_2, \ldots, \mathbf{N}_t\}$ are $m \times n$ random matrices with i.i.d entries drawn from a zero-mean circular symmetric complex Gaussian distribution with standard deviation $\frac{\sigma}{\sqrt{t}}$ where $t$ indicates the maximum number of colluding workers for privacy. Let $\gamma = e^{-\frac{2\pi \iota}{N}}$ and $\omega = e^{-\frac{2\pi \iota}{k+t}}$ be the $N$-th and $(k+t)$-th roots of unity, respectively, with $\iota^2 =-1$. With the above ingredients, the Lagrange polynomial is constructed as 
 \[ u(z)=\sum_{r=1}^{k} \mathbf{X}_r l_r(z) + \sum_{r=k+1}^{k+t} \mathbf{N}_{r-k}l_{r}(z) =\sum_{r=1}^{k+t} \mathbf{Z}_r l_r(z),\] 
where $l_r(.)$'s are Lagrange monomials defined as
\begin{equation*}
l_r(z)=\prod_{ l\in[k+t] \backslash r} \frac{z-\beta_l}{\beta_r-\beta_l},
\end{equation*}
for all $r\in[k+t]$. Note that $\beta_r$'s are picked to be equally spaced on a circle of radius $\beta$ centered around 0 in the complex plane such that $\beta_r =\beta \omega^{r-1}$ for $\beta \in \mathbb{R}$. Finally, the shares of the encoded data set to be distributed among the workers involves evaluation of $u(z)$ over the $N$-th roots of unity in the complex plane i.e., $\mathbf{Y}_i = u(\alpha_i)$ such that  $\alpha_i=\gamma^{i-1}$, is sent to the $i$-th worker, for $i\in[N]$.

\subsection{Decoding in ALCC}
After receiving its share from the master node, the $i$-th node computes $f(\mathbf{Y}_{i}) \in \mathbb{R}^{u \times h}$ and returns the result to the master node. The master node then interpolates the polynomial $f(u(z))$ by using the results returned from at least $(k+t-1)D+1$ workers. Note that this is the minimum number of returned evaluations needed to guarantee a successful interpolation of $f(u(z))$ since $f(u(z))$ has degree $(k+t-1)D$. If $\tilde D$ is used to denote $(k+t-1)D$, which indicates the degree of the polynomial $f(u(z))$, then the number of workers required for successful interpolation of $f(u(z)$ is $K = \tilde D +1$. Therefore, the  maximum number of stragglers that can be tolerated by ALCC is $N-K$. Finally, to recover $f(\mathbf{X}_{i})$’s, the master node computes $f(u(\beta_r))$ for $r\in[k]$.

As each stage of ALCC involves floating-point operations, accuracy becomes a crucial metric for evaluation. For a given data set $\mathbf{X}$, let $\mathbf{Y'} = f(\mathbf{X})$ denote the computation using ALCC through floating point operations. Similarly, let $\mathbf{Y} = f(\mathbf{X})$ denote the computation at the master node without using ALCC through floating point operations. Using $\mathbf{Y}$ and $\mathbf{Y'}$, the relative error introduced by ALCC with respect to the centralized computation is captured by $e_{rel} \triangleq \frac{||\mathbf{Y}-\mathbf{Y'}||}{||\mathbf{Y}||},$
 where $||.||$ denotes the $l^2$-norm. 
 
\section{ALCC with Adversarial Workers}
\label{sec:ALCC_adv}
In contrast to the vanilla ALCC, we consider a setting wherein $A$ out of the $N$ workers send erroneous computation results to the master node. To keep the focus solely on adversarial workers, throughout the paper, we assume the absence of stragglers and honest-but-curious workers. However, all the results of this work can be generalized to the case of stragglers and honest-but-curious workers. Formally, let $i_{1}, i_{2}, \ldots, i_{A}$ be the indices of the adversarial workers, wherein the adversary $i_{a}$ is expected to return $f(\textbf{Y}_{i_{a}})$ for each $a \in [A]$. Incorporating the additive noisy adversarial model, if node $i_{a}$ is an adversarial worker, we assume that it returns $\textbf{R}_{i_{a}}= f(\textbf{Y}_{i_{a}})+\mathbf{E}_{i_{a}} \in \mathbb{R}^{u \times h}$ where $\mathbf{E}_{i_{a}} \in \mathbb{R}^{u \times h}$ is the noise matrix chosen by the adversary. As a result of the error introduced by the adversaries, the subsequent steps of interpolation and evaluation gets effected, thereby resulting in large relative errors. 

Towards solving the problem of ALCC with adversarial workers, let $\{\mathbf{R}_{i} \in \mathbb{R}^{u \times h} ~|~ i \in [N]\}$ denote the set of computations returned by the workers such that $\mathbf{R}_{i} = f(\mathbf{Y}_{i}) + \mathbf{E}_{i}$ if $i = i_{a}$ for some $a \in [A]$, otherwise, $\mathbf{R}_{i} = f(\mathbf{Y}_{i})$. The following proposition is crucial for the rest of the contributions. 

\begin{proposition}
\cite{ShH} For an ALCC setting with parameters N and $K = (k+t-1)D + 1$, the erroneous computations returned by the $A$ adversarial workers can be nullified as long as $A \leq v \triangleq \lfloor \frac{N-K}{2}\rfloor$ and the floating-point operations of ALCC have infinite precision.  
\end{proposition}
\begin{IEEEproof}
For every $\bar{u} \in [u], \bar{h} \in [h]$, let $\mathbf{r}_{\bar{u}\bar{h}} = [\mathbf{R}_{1}(\bar{u}, \bar{h}) ~\mathbf{R}_{2}(\bar{u}, \bar{h}) ~\ldots~ \mathbf{R}_{N}(\bar{u}, \bar{h})]$, represent a vector carved from the $(\bar{u}, \bar{h})$-th entry of each $\mathbf{R}_{i}$. With $A = 0$, $\mathbf{r}_{\bar{u}\bar{h}}$ corresponds to a codeword of a $K$-dimensional Discrete Fourier Transform (DFT) code with blocklength $N$. This observation follows from the fact the generator matrix of an $(N,K)$ DFT code consists of any $K$ columns from the DFT matrix of order $N$ \cite{b7}, and moreover, in the encoding stage of ALCC, the encoded shares are evaluated at the $N$-th roots of unity. Thus, every evaluation of $f(u(z))$ at the $N$-th roots of unity, can be viewed as a codeword of DFT code. In addition, since $(N,K)$ DFT codes have error-correction capability of at most $\lfloor \frac{N-K}{2}\rfloor$ errors, their corresponding error-correction algorithms can be applied on each $\mathbf{r}_{\bar{u}\bar{h}}$. Finally, since the error vectors are real-valued, infinite precision on floating point representation is necessary to perfectly localize the errors and cancel them. 
\end{IEEEproof}

For DFT codes, both coding-theoretic and subspace-based error-correction algorithms are popular \cite{b3},\cite{b5},\cite{b6}, \cite{b9}, \cite{b10}. Broadly, their implementation involves the steps of (i) calculation of syndrome vector, (ii) estimation of the number of errors, (iii) localization of errors, and (iv) estimation of error values. As a result, these algorithms can be independently applied on each noisy DFT codeword $\mathbf{r}_{\bar{u}\bar{h}}$ to recover an estimate of $\{f(\mathbf{Y}_{i}), i \in [N]\}$. Although classical DFT decoders are effective in infinite precision environments, they suffer performance degradation with finite precision floating point operations \cite{b8}.

\subsection{On Independent Error Localization using DFT Decoder}
\label{DFT_decoder_basics}

To capture finite precision operations at the workers, let the matrices recovered at the master node be $\mathbf{R}_{i} = f(\mathbf{Y}_{i}) + \mathbf{E}_{i} + \mathbf{P}_{i}$ if $i = i_{a}$ for some $a \in [A]$, otherwise, $\mathbf{R}_{i} = f(\mathbf{Y}_{i}) + \mathbf{P}_{i}$, where $\mathbf{P}_{i} \in \mathbb{R}^{u \times h}$ captures the precision errors. To study the effect of precision errors, assume that the master node recovers a total of $M = u \times h$ noisy DFT codewords and decides to apply a DFT decoder on each codeword \emph{independently}. While all the internal blocks of a DFT decoder suffer from degradation due to precision errors in $\{\mathbf{R}_{i}\}$, the most vulnerable block is the step of error localization. Towards detecting the error locations corresponding to the vector $\mathbf{r}_{\bar{u}\bar{h}}$, we assume that the number of errors has been correctly estimated as $A$. Further, let $\bar{g}_{\bar{u}\bar{f}}(x)$ denote the error-locator polynomial of degree $A$ given by
$\bar{g}_{\bar{u}\bar{f}}(x) = g_{\bar{u}\bar{f}}(x) + e_{\bar{u}\bar{f}}(x),$ where $g_{\bar{u}\bar{f}}(x)$ denotes the actual error-locator polynomial computed using the syndrome vector with infinite precision and $e_{\bar{u}\bar{f}}(x)$ is the error polynomial due to precision errors. Let the coefficients of $e_{\bar{u}\bar{f}}(x)$ be i.i.d. as $\mathcal{CN}(0, \sigma^{2}_{p})$. With infinite precision, the roots of the error-locator polynomial accurately provides the positions of the errors. However, with precision errors, one needs to evaluate $||\bar{g}_{\bar{u}\bar{f}}(X_{q})||^2$, where $X_{q} = e^{-\iota 2 \pi q/N}$ for each $q\in[N]$ and arrange the evaluations in the ascending order in order to obtain $$||\bar{g}_{\bar{u}\bar{f}}(X_{\hat{i}_{1}})||^2 \leq||\bar{g}_{\bar{u}\bar{f}}(X_{\hat{i}_{2}})||^2\leq \ldots \leq||\bar{g}_{\bar{u}\bar{f}}(X_{\hat{i}_{N}})||^2,$$ where \{$\hat{i}_1,\hat{i}_2, \ldots, \hat{i}_N\}$ is the ordered set based on the evaluations. Subsequently, the $A$ smallest evaluations and its corresponding set $\hat{\mathcal{L}} = \{\hat{i}_1,\hat{i}_2, \ldots, \hat{i}_A\}$ are treated as the detected error locations.

Given that the true locations of errors are $\mathcal{L} = \{i_{1}, i_{2}, \ldots, i_{A}\}$, the error rate of localization can be captured as $P_{error} = \sum_{a= 1}^{A} P(i_{a} \notin \hat{\mathcal{L}})$. Furthermore, for $i_{a}$ to not appear in $\hat{\mathcal{L}}$, there should be another index $i_{b} \in [N] \backslash \mathcal{L}$ such that $||g_{\bar{u}\bar{f}}(X_{i_{b}})||^2 < ||g_{\bar{u}\bar{f}}(X_{i_{a}})||^2$. While $i_{b}$ can take several values, we only consider the dominant event when $i_{b}$ is the nearest neighbour of $i_{a}$, i..e, when $i_{b} = i_{a} + 1$ modulo $N$ or $i_{b} = i_{a} - 1$ modulo $N$. Note that these events are relevant when the variance of the precision error is small, thus resulting in $P_{error} \approx 2A \times PEP_{dom}$, where $PEP_{dom}$ is the pairwise error probability between the neighbouring indices $i_{a}$ and $i_{b}$, given by
\begin{equation}
\label{eq:PEP}
PEP_{dom} = P(||g_{\bar{u}\bar{f}}(X_{i_{a}})||^2\leq ||g_{\bar{u}\bar{f}}(X_{i_{b}})||^2),
\end{equation}
where $P(\cdot)$ is the probability operator.

Since closed-form expressions on $PEP_{dom}$ are intractable to obtain, we propose a lower bound on it. 

\begin{theorem}
\label{thm:corr}
For any $A, N$ and $\sigma^{2}_{p}$, a lower bound on the pairwise error probability in \eqref{eq:PEP} can be expressed as
\begin{equation}
\label{eq:lower_bound}
PEP_{dom} \geq PEP_{L} \triangleq \frac{\kappa}{(1+\kappa)}e^{-\frac{\eta (c_I^2 +c_Q^2)\kappa}{4\sigma^2_p(1+\kappa)}},
\end{equation}
where $\eta, c_I$, $c_{Q}$ are constants, $\kappa =\frac{2}{\eta \sum_{j = 1}^{A} 1 - \mbox{cos}(\frac{j2\pi}{N})}$.
\end{theorem}

Based on Theorem \ref{thm:corr}, we present the following corollary.

\begin{corollary}
\label{lemma:sigma_beh}
For a given $A$ such that $A \leq \lfloor \frac{N-K}{2}\rfloor$, the lower bound in Theorem \ref{thm:corr} is a non-decreasing function of $\sigma^2_p$.
\end{corollary}

Given that a lower bound on $PEP_{dom}$ increases with increasing $\sigma^{2}_{p}$, the error rate of localization is expected to degrade for non-negligible variance of precision errors, thereby rendering the DFT decoder ineffective when used as an off-the-shelf block. In the next section, we show that the error-localization block of DFT decoders can be customized to assist ALCC in improving its end-to-end accuracy. 




\subsection{On Joint Error Localization using DFT Decoder}

At the input of the error-localization block of DFT decoder, a total of $M = u \times h$ error-locator polynomials are available along with the information on their degrees. Since the sparsity of the noise matrices $\{\mathbf{E}_{i_{a}}, a \in [A]\}$ can be arbitrary and is unknown to the master node, the degrees of the $M$ error-locator polynomials can be arbitrary. However, in the special case when all the error-locator polynomials have degree $v = \lfloor \frac{N-K}{2}\rfloor$, the following proposition shows that the master node can localize the errors by evaluating the roots of only one error-locator polynomial, which is obtained by component-wise averaging of all the coefficients of the $M$ polynomials.  

\begin{proposition}
\label{prop:all_one_matrix}
When $\sigma^{2}_{p} > 0$, if $\mathbf{E}_{i_{a}}$ has non-zero entries in all its locations, for every $a \in [v]$, then the ALCC based localization can reduce its error rate compared to that in independent localization.   
\end{proposition}

In general, when the degrees of the $M$ error-locator polynomials are arbitrary, the master node can identify all the degree-$v$ polynomials and average them to compute one polynomial from them. However, the rest of the polynomials that have degree less than $v$ cannot be averaged since their error locations are not guaranteed to be identical. Furthermore, since there are at most $v$ adversarial workers, the cardinality of the union of the roots of all the $M$ polynomials should be bounded by $v$. Thus, instead of solving for the roots of each polynomial independently, we emphasize the need for jointly solving the roots of all the polynomials.

Given the error-locator polynomials $\{\bar{g}_{\bar{u}\bar{f}}(x), \bar{u} \in [u], \bar{f} \in [f]\}$, let $\mathcal{G}_{v}$ and $\mathcal{G}_{v^{-}}$ denote the set of polynomials with degree $v$ and degree less than $v$, respectively. Then, we obtain a new set of polynomials, denoted by $\mathcal{G}_{joint} = \{\bar{g}_{max}\} \cup \mathcal{G}_{v^{-}}$, where $\bar{g}_{max} = \frac{1}{|\mathcal{G}_{v}|} \sum_{\bar{g}_{\bar{u}\bar{f}}(x) \in |\mathcal{G}_{v}|} \bar{g}_{\bar{u}\bar{f}}(x)$ is the polynomial obtained by averaging the degree $v$ polynomials. Furthermore, let $\bar{M} \leq M$ represent the cardinality of $\mathcal{G}_{joint}$. We now propose a method to jointly solve for the roots of $\mathcal{G}_{joint}$. Assuming that the polynomials of $\mathcal{G}_{joint}$ are sorted in some order, let $t_{m}(x) \in \mathcal{G}_{joint}$ be the $m$-th polynomial of degree $d(m)$, where $1 \leq d(m) \leq v$. We propose to solve for the roots of every polynomial in $\mathcal{G}_{joint}$ independently, and then take the union of the roots. Let this set of roots be denoted by $\mathcal{S}_{I}$, which is of cardinality $v''$. Furthermore, let $\mathcal{S}_{total} = \mathcal{S}_{I}^{v}$ be the set of all $v$-tuples over the set $\mathcal{S}_{I}$. For a given $\mathcal{S} \in \mathcal{S}_{total}$, we evaluate $t_{m}(x)$ at all the entries of $\mathcal{S}$ as explained in Section \ref{DFT_decoder_basics}, and obtain its first $d(m)$ smallest evaluations. Let the sum of the first $d(m)$ smallest evaluations of $t_{m}(x)$ be denoted by $E_{t_{m}(x)}(\mathcal{S})$, and the corresponding roots be $R_{t_{m}(x)}(\mathcal{S}) \subset \mathcal{S}$. Thus, the sum of the first $d(m)$ smallest evaluations of polynomials over the entire set $\mathcal{G}_{joint}$ corresponding to $\mathcal{S}$ is $\mathcal{E}_{\mathcal{S}} \triangleq \sum_{m = 1}^{\bar{M}} E_{t_{m}(x)}(\mathcal{S})$. Also, the set of the roots of all the polynomials in $\mathcal{G}_{joint}$ corresponding to $\mathcal{S}$ is denoted by $\mathcal{R}_{\mathcal{S}} \triangleq \{R_{t_{1}(x)}(\mathcal{S}), R_{t_{2}(x)}(\mathcal{S}), \ldots, R_{t_{\bar{M}}(x)}(\mathcal{S})\}$. Having obtained $\mathcal{E}_{\mathcal{S}}$ and $\mathcal{R}_{\mathcal{S}}$, we propose to solve 
\begin{equation}
\label{opt_Problem}
\hat{\mathcal{S}} = \arg \min_{\mathcal{S} \in \mathcal{S}_{total}} \mathcal{E}_{\mathcal{S}},
\end{equation}
and then obtain $\mathcal{R}_{\hat{\mathcal{S}}}$ as its solution. 

Note that as $v''$ (the cardinality of $\mathcal{S}_{I}$) increases, the computational complexity to solve \eqref{opt_Problem} also increases. However, in practice, if $\sigma^{2}_{p}$ is small, then even if $v'' \geq v$, $v'' - v$ is expected to be small, thereby facilitating a solution to \eqref{opt_Problem} with computational complexity of the order of $v'' \choose v$. To further reduce the complexity, we can replace $\mathcal{S}_{I}$ by $\mathcal{S}_{II}$, where $\mathcal{S}_{II} \subseteq \mathcal{S}_{I}$ is chosen at random such that the cardinality of $\mathcal{S}_{II}$ is $v' \leq v''$. Henceforth, $v'$ is referred to as the set constraint length for the proposed joint localization step, and this parameter can be chosen to trade the computational complexity with the error rate of joint localization.

\begin{figure}[ht!]
\includegraphics[scale = 0.29]{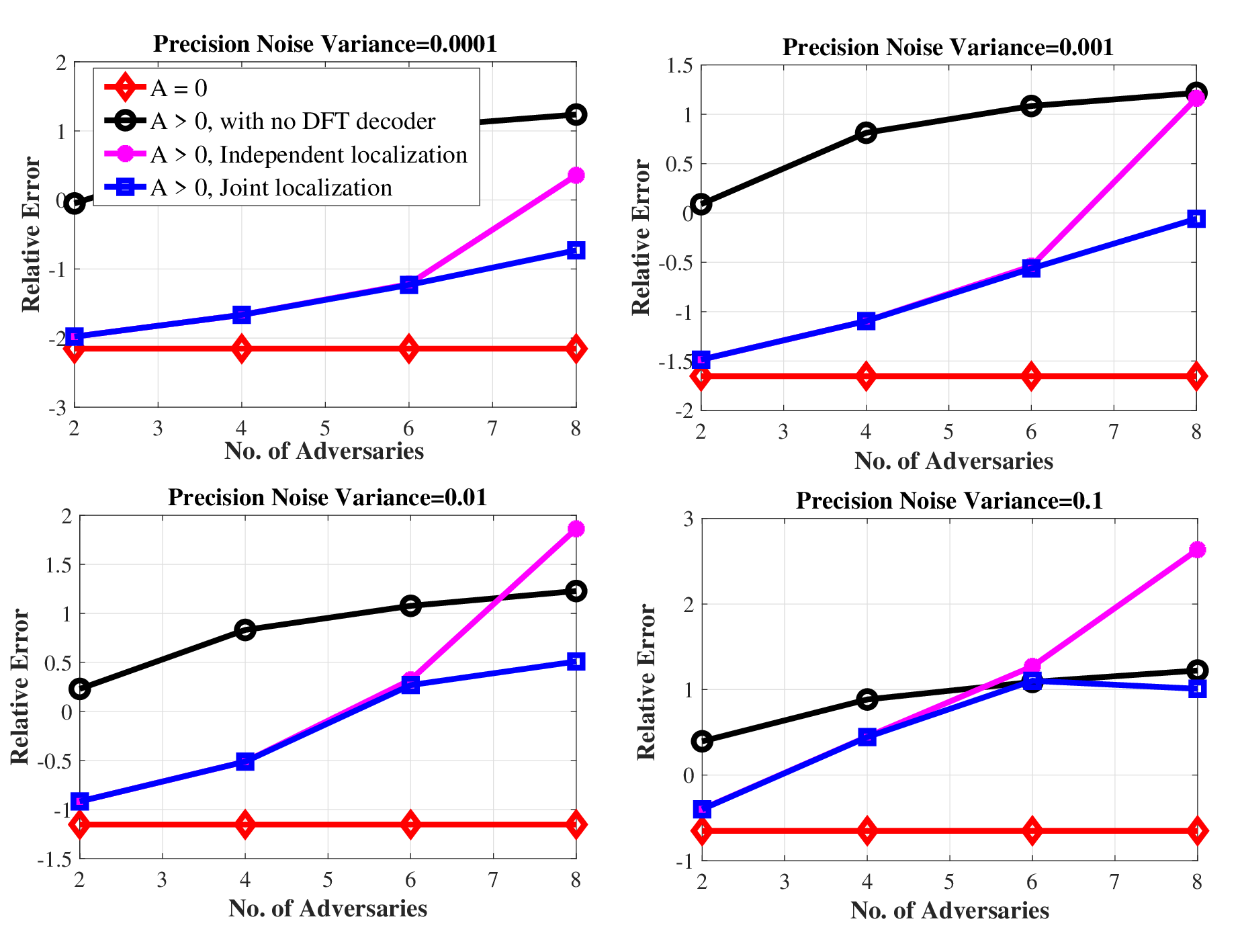}
\caption{Average relative error (in dB scale) of ALCC with parameters $N = 31, K = 15, \beta = 1.5, t = 3, \sigma = 10^6$ with and without the DFT decoders in the presence of adversarial workers. Here, the non-zero entries of adversarial matrices $\{\mathbf{E}_{i_{a}}\}$ are i.i.d as $\mathcal{CN}(10, 10^3)$.}
\label{joint2}
\end{figure}

To demonstrate the benefits of joint error localization as part of DFT decoders, we present experimental results on the average accuracy of ALCC in Fig. \ref{joint2} for various values of $A$ when computing $f(\mathbf{X}) = \mathbf{X}^{T}\mathbf{X}$ such that $\mathbf{X} \in \mathbb{R}^{20 \times 5}$. To generate the results on joint localization, $\mathcal{S}_{total} = \mathcal{S}_{I}^{v}$ was used. The plots confirm that joint localization outperforms independent localization especially when $A$ is close to $v$.

\section{Optimized Colluding Attacks on ALCC}
\label{sec:colluding_attacks}

With $i_{1}, i_{2}, \ldots, i_{v}$ denoting the indexes of the adversarial nodes, such that $v = \lfloor \frac{N-K}{2}\rfloor$, we assume that these adversaries can collude through a separate private channel to arrive at an attack strategy. Although collusion may be needed to decide the choice of the noise matrices $\mathbf{E}_{i_{1}}, \mathbf{E}_{i_{2}}, \ldots, \mathbf{E}_{i_{v}}$, we limit our discussion for the setting when the adversaries collude to decide on the sparsity of their noise matrices. Once the sparsity and the positions of the non-zero entries are chosen, we assume that non-zero entries are injected on those positions from a suitable underlying distribution in a statistically independent manner. Formally, for each $\mathbf{E}_{i_{a}}$, such that $a \in [v]$, let $\mathbf{B}_{i_{a}} \in \{0, 1\}^{u \times h}$ represent the binary matrix that captures the positions of the non-zero entries of $\mathbf{E}_{i_{a}}$. Henceforth, these matrices are referred to as the \emph{base matrices} of the adversaries. For instance, if $\mathbf{B}_{i_{a}} = \mathcal{J}_{u \times h}$, where $\mathcal{J}_{u \times h}$ denotes the all-one matrix of size $u \times h$, this implies that every component of the returned matrix $f(\mathbf{Y}_{i_{a}})$ is corrupted with an additive noise by the worker $i_{a}$. Before introducing the problem statement on the choice of the matrices $\{\mathbf{B}_{i_{1}}, \mathbf{B}_{i_{2}}, \ldots, \mathbf{B}_{i_{v}}\}$, Proposition \ref{prop:all_one_matrix} reminds us on how not to choose them. 

Given that the choice of $\mathbf{B}_{i_{a}} = \mathcal{J}_{u \times h}$, $\forall a \in [v]$ helps the master node to reduce the error rates, the adversaries would need to choose their base matrices such that the master node should not average out the error locator polynomials. To define a problem statement on choice of the base matrices $\{\mathbf{B}_{i_{1}}, \mathbf{B}_{i_{2}}, \ldots, \mathbf{B}_{i_{v}}\}$, we define the effective base matrix, denoted by $\mathbf{B}_{eff} \in \{0, 1\}^{uh \times v}$, as 

\begin{small}
\begin{equation}
\mathbf{B}_{eff} = \begin{bmatrix}
\mathbf{B}_{i_{1}}(1,1) & \mathbf{B}_{i_{2}}(1, 1) & \ldots & \mathbf{B}_{i_{v}}(1, 1)\\
\mathbf{B}_{i_{1}}(1,2) & \mathbf{B}_{i_{2}}(1, 2) & \ldots & \mathbf{B}_{i_{v}}(1, 2)\\
\vdots & \vdots & \vdots & \vdots\\
\mathbf{B}_{i_{1}}(1,h) & \mathbf{B}_{i_{2}}(1, h) & \ldots & \mathbf{B}_{i_{v}}(1, h)\\
\mathbf{B}_{i_{1}}(2,1) & \mathbf{B}_{i_{2}}(2, 1) & \ldots & \mathbf{B}_{i_{v}}(2, 1)\\
\vdots & \vdots & \vdots & \vdots\\
\mathbf{B}_{i_{1}}(u,h) & \mathbf{B}_{i_{2}}(u, h) & \ldots & \mathbf{B}_{i_{v}}(u, h)\\
\end{bmatrix},	
\end{equation}
\end{small}

\noindent where $\mathbf{B}_{i_{a}}(\bar{u},\bar{h})$ denotes the $(\bar{u},\bar{h})$-th entry of $\mathbf{B}_{i_{a}}$. 

The above representation is such that the rows of $\mathbf{B}_{eff}$ represent the $M$ error locator polynomials, and the number of ones in a given row captures the degree of the corresponding error locator polynomial. Note that, for $1 \leq a \leq v$, the $a$-th column of $\mathbf{B}_{eff}$ is completely in control of the $a$-th adversary, but not the elements across the columns. As a result, the adversaries need to collude in order to control the number of ones in a row, which in turn controls the degree of the error locator polynomials. With this background, an interesting problem from the perspective of the adversaries is to choose $\mathbf{B}_{eff}$ such that the error rate of the localization step is maximized. However, given the combinatorial nature of the optimization problem in \eqref{opt_Problem}, it is intractable to obtain exact expressions on the error rate of joint localization for various set constraint lengths. 
Therefore, to target an analytically-tractable variant of the joint localization step, we show that the adversaries can choose $\mathbf{B}_{eff}$ to maximize the error rate of a joint localization method, wherein independent localization is applied on $\mathcal{G}_{joint}$, after averaging all the degree $v$ polynomials. Towards that direction, we use the following relation between $A$ and the lower bound on the error rate to choose $\mathbf{B}_{eff}$. 


\begin{lemma}
\label{lemma:inc_A}
For a given $N$, when $\sigma^{2}_{p}$ is small, the lower bound $A \times PEP_{L}$ is a non-decreasing function of $A$.
\end{lemma}

To propose an appropriate choice of $\mathbf{B}_{eff}$, we use Lemma \ref{lemma:inc_A} and Proposition \ref{prop:all_one_matrix} to define two classes of colluding attacks, namely the strongly colluding and weakly colluding attacks.

\subsection{Strongly Colluding Attack}

In this threat model, we assume that the $v$ adversaries exchange messages in order to jointly arrive at $\mathbf{B}_{eff}$. In particular, we assume that one of the $v$ adversaries is a master adversary, which has the knowledge of all the underlying parameters of ALCC. Subsequently, it chooses $\mathbf{B}_{eff}$ and then distributes it to the other adversaries. Given that the size of $\mathbf{B}_{eff}$ can be large, this model requires high communication-overhead in the colluding channels. The following proposition proposes the optimal way to choose $\mathbf{B}_{eff}$. 

\begin{proposition}
\label{prop:strog_coll}
In the strongly colluding attack, when $\sigma^{2}$ is small, the optimal way to choose $\mathbf{B}_{eff}$ is to pick the all-one row as its first row, and then for each of the other rows, pick $v-1$ ones at random positions with uniform distribution.
\end{proposition}

While the master node cannot average out the precision noise against the above attack, it may solve \eqref{opt_Problem} using the constraint lengths of its choice to further improve the accuracy. 

\subsection{Weakly Colluding Attack}

When the private channel between the adversaries cannot support high communication-overhead, the master adversary may not explicitly communicate $\mathbf{B}_{eff}$. To circumvent this problem, we propose a threat model, wherein the master node sends a \emph{common seed} to the adversaries, using which they generate their column of $\mathbf{B}_{eff}$ in a probabilistic manner. In particular, the master adversary uses the knowledge of the ALCC parameters to choose a probability value $p \in (0, 1)$. Subsequently, a suitably quantized version of $p$ is broadcast depending on the communication-overhead restrictions. Then, each adversary will populate its column in $\mathbf{B}_{eff}$ by generating its entries in a statistically independent manner using a Bernoulli process with probability $p$. Since identical $p$ is used at all the adversaries, this process is equivalent to generating $\mathbf{B}_{eff} \sim \mbox{Bernoulli}(p)$. Henceforth, let $p$ denote the probability associated with bit 0 when using $\mbox{Bernoulli}(p)$. While this threat model requires less communication-overhead over the strongly colluding model, it may not generate $\mathbf{B}_{eff}$ with the structure of Proposition \ref{prop:strog_coll} at each instant, thus justifying its name. The following proposition presents a way to choose $p$.

\begin{proposition}
\label{prop_weak}
 When $\sigma^{2}_{p}$ is small, a reasonable strategy to pick $p$ under the weakly colluding model is to solve
\begin{equation}
\label{eq:pval}
p^{*} = \arg \min_{\ p \in (0, 1)} \frac{1}{Mv}\mathbb{E}[N_{0}] + \frac{1}{M}\mathbb{E}[N_{\textbf{1}}],
\end{equation}
where $N_{0}$ denotes the number of zero entries in $\mathbf{B}_{eff}$ and $N_{\textbf{1}}$ denotes the number of rows in $\mathbf{B}_{eff}$ that have all ones.
\end{proposition}
\begin{IEEEproof}
When $\sigma^{2}_{p}$ is small, the lower bound on error rates in localization is an increasing function of $A$ in the range $[v]$. As a consequence, $\mathbf{B}_{eff}$ must be such that the average number of zeros (captured by $\mathbb{E}[N_{0}] = Mvp$) in it should be minimized. Also, given that the master node can perform averaging of all the degree $v$ polynomials, the average number of rows that have all ones in it, denoted by $\mathbb{E}[N_{\textbf{1}}] = M(1-p)^{v}$, should be minimized. Thus, combining the two objectives, we propose $p^{*}$ in \eqref{eq:pval}, which can be obtained in closed-form.
\end{IEEEproof}

\begin{figure}[ht!]
\includegraphics[scale = 0.31]{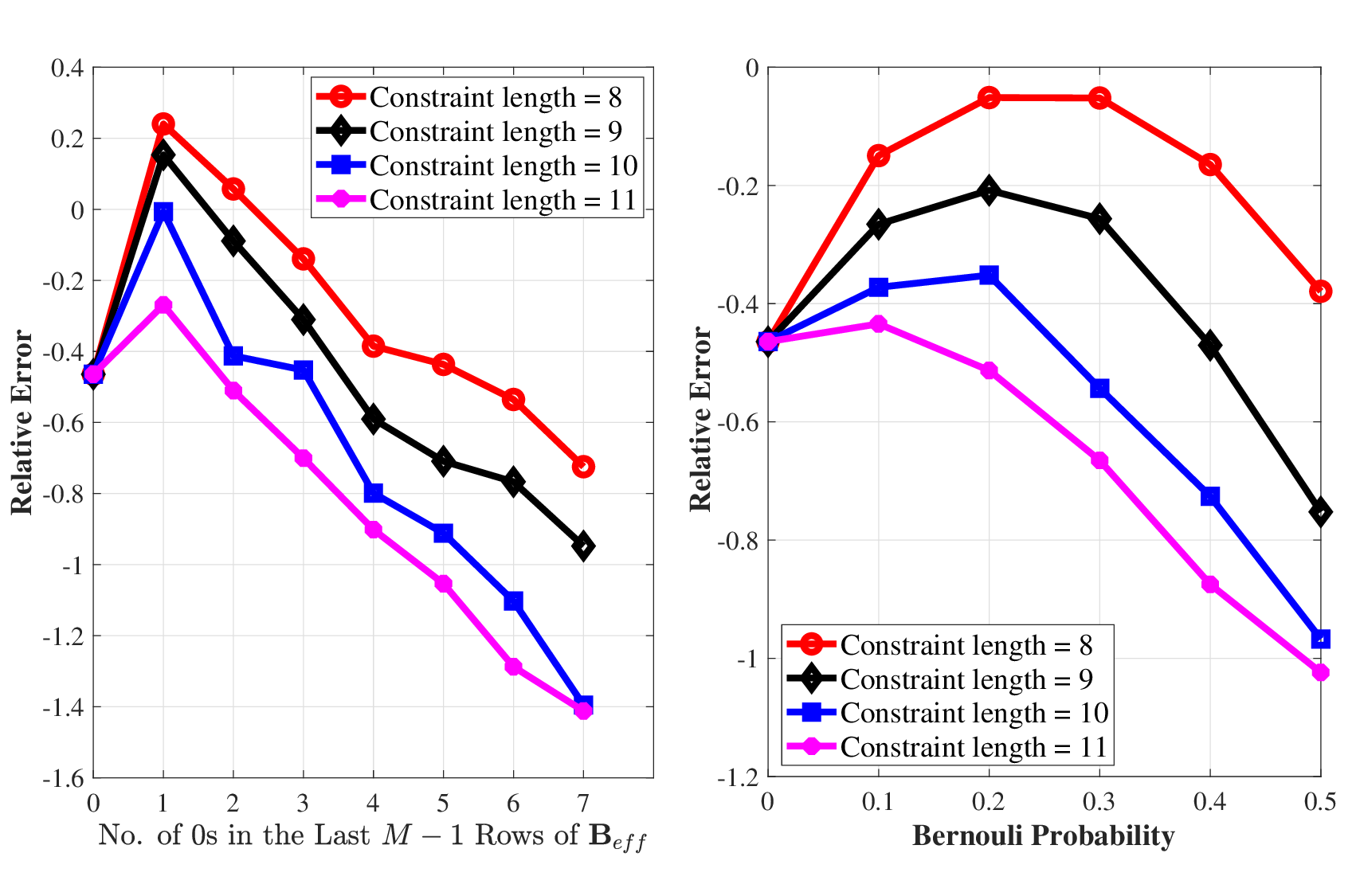}
\vspace{-0.65cm}
\caption{Average relative error (in dB scale) of ALCC against strongly and weakly colluding attacks for $N = 31, K = 15$, $\sigma^{2}_{p} = 0.01$. For the strongly colluding case, the first row of $\mathbf{B}_{eff}$ has all ones, whereas the other rows have variable number of ones. For the weakly-colluding case, using $p^{*} = 0.257$ provides the worst-case accuracy for constraint length 8 as pointed above.}
\label{fig:colluding}
\end{figure}
\hspace{-5mm}

\vspace{-0.65cm}

To showcase the benefit of both the colluding attacks, in Fig. \ref{fig:colluding}, we present experimental results on the average relative error of ALCC when joint localization with varying constraint lengths are used. To generate the plots, we assume that the master node intends to compute $f(\mathbf{X}) = \mathbf{X}^{T}\mathbf{X}$ such that $\mathbf{X} \in \mathbb{R}^{20 \times 5}$. Furthermore, for the attack model, the non-zero entries of adversarial matrices $\{\mathbf{E}_{i_{a}}\}$ are i.i.d as $\mathcal{CN}(10, 10^2)$. The plots confirm that the adversaries can use the results of Proposition \ref{prop:strog_coll} and Proposition \ref{prop_weak} to design their base matrices when $\sigma^{2}_{p} > 0$. From the viewpoint of the master node, the plots indicate that the only way to alleviate the impact of the proposed attacks is to increase the computational complexity, which may not be desirable.





$~~$\\
\newpage

\end{document}